\documentclass{ws-procs9x6}
\def\la{\langle}\def\ra{\rangle}
\def\bea{\begin{eqnarray}}
\def\eea{\end{eqnarray}}
\def\lsim{\mathrel{\rlap{\lower3pt\hbox{\hskip1pt$\sim$}}
     \raise1pt\hbox{$<$}}} 
\def\gsim{\mathrel{\rlap{\lower3pt\hbox{\hskip1pt$\sim$}}
     \raise1pt\hbox{$>$}}} 

\def\chisu3{SU(3)_L\times SU(3)_R}
\def\calL{\mathcal L}
\def\tr{\rm tr}
\def\np{Nucl. Phys.}\def\prl{Phys. Rev. Lett.}\def\pr{Phys. Rev.}
\def\calH{\mathcal H}
\def\del{\partial}

\begin{document}

\title{Chiral Symmetry of Heavy-Light-Quark Hadrons in Hot/Dense Matter
\footnote{\uppercase{T}he concluding talk at the
\uppercase{KIAS}-\uppercase{APCTP} \uppercase{S}ymposium in
\uppercase{A}stro-\uppercase{H}adron \uppercase{P}hysics
``\uppercase{C}ompact \uppercase{S}tars: \uppercase{Q}uest for
\uppercase{N}ew \uppercase{S}tates of \uppercase{D}ense
\uppercase{M}atter," \uppercase{N}ovember 10-14, 2003,
\uppercase{S}eoul, \uppercase{K}orea.}}

\author{Mannque Rho
}

\address{Service de Physique Th\'eorique, CEA/Saclay, 91191 Gif-sur-Yvette,
France,\\ Korea Institute for Advanced Study, Seoul
130-722, Korea \\
\& Department of Physics, Hanyang University, Seoul 133-791, Korea\\
E-mail: rho@spht.saclay.cea.fr}

\maketitle

\abstracts{The recent discoveries by the BaBar and CLEO II
collaborations on the splitting between $D_s$ and ${\tilde D}_s$
which exhibited surprises in the structure of heavy-light-quark
systems are connected -- via the Harada-Yamawaki ``vector
manifestation" of hidden local symmetry -- to chiral symmetry
restoration expected to take place at some critical temperature
$T_c$ in heavy-ion collisions or at some critical density $n_c$ in
the deep interior of compact stars, the main theme of this
symposium. This unexpected connection exemplifies the diversity of
astro-hadronic phenomena discussed in this meeting.}

\section{Foreword}
\indent\indent This is the last talk of this Symposium and as such
it is supposed to conclude it. What distinguishes this meeting
from other meetings of a similar scope is the diversity of the
topics covered, ranging from hadron/particle physics to
astrophysics, but with a common objective, that is, to explore the
extreme state of matter in high density or/and high temperature.
Given the diversity and the exploratory stage of the development,
it would be presumptuous of me to make any conclusion on any
subtopic of the meeting, not to mention the totality, so what I
will do is to present to you yet another surprising development
that at first sight might appear unrelated to the main theme of
the meeting but as it turns out, has an uncanny connection to what
we have been discussing throughout this meeting.

What I shall present here is a combination of the work I did with
Maciej A. Nowak and Ismail Zahed a decade ago~\cite{NRZ93} and the
work I did very recently with Masayasu Harada and Chihiro
Sasaki~\cite{HRS-babar}. The question I will address here is: {\it
Do the recent discoveries on the structure of heavy-light-quark
mesons by the BaBar, CLEO and Belle collaborations that we have
just heard have any ramifications on what we have been discussing
in this meeting, i.e., the structure of compact stars and the
early Universe?}\ My answer is: {\it Yes, if the chiral phase
transition predicted in QCD has significant influence on them.}
\section{The BaBar, CLEO and Belle Discoveries}
\indent\indent On April 12th 2003, the BaBar collaboration
announced a narrow peak of mass 2.317GeV $/c^2$ that decays into
$D_s^+\pi^0$~\cite{BABAR}. On May 12th 2003, the CLEO II
collaboration confirmed the BaBar result, and also observed a
second narrow peak of mass 2.46 GeV$/c^2$ in the final
$D_s^{*+}\pi^0$ state~\cite{CLEO}. Subsequently both states were
confirmed by the Belle collaboration~\cite{belle}. The
experimental results were surprising since such states were
expected neither to lie below $DK$ and $D^*K$ thresholds nor to be
so narrow. These observations have recently generated a flurry of
theoretical activities. The excitement surrounding these
developments was nicely summarized by Nowak~\cite{nowak}.

Remarkably, however, the existence of this type of states was
theoretically predicted more than a decade ago~\cite{NRZ93,BH94}
based on the combination of chiral symmetry of light quarks and
heavy-quark symmetry of heavy quarks. What is relevant in my
discussion in this meeting is the suggestion made in \cite{NRZ03}
that the splitting of the chiral doublers carries a direct
information on the light-quark condensate $\la\bar{q}q\ra$ and can
therefore be a litmus indicator for chiral symmetry property of
the medium in which the chiral doubling phenomenon is observed. In
particular, if one measures the splitting in hot or dense matter,
then as the chiral phase transition point generically denoted
$[pt]_\chi$ (such as the critical temperature $T_c$ or density
$n_c$) is approached, the splitting should disappear in the chiral
limit. This could then be an ideal tool to map out the chiral
phase structure of hot/dense matter.
\section{Chiral Doubling Starting From the ``Vector Manifestation (VM)"}
 \indent\indent The standard approach to
hot/dense-matter physics starts with a Lagrangian for cold/dilute
matter for which one has both experimental and theoretical control
and then drives the system to a hot/dense environment so as to
bring it to a phase transition. This is what is being done in
heavy-ion collisions and in compact-star physics heard in this
meeting. This was the idea proposed by Nowak, Rho and Zahed in
\cite{NRZ03} for heavy-light mesons in medium. The idea of Harada,
Sasaki and myself~\cite{HRS-babar} is to go in the opposite
direction: Instead of going from zero temperature/density to high
temperature/density, that is, ``bottom-up", we will go
``top-down". This is because we think we have a theory which is
well-defined at the critical point $[pt]_\chi$ although whether
that theory is directly related to QCD is yet to be verified. Our
task then is to simply assume that this description of $[pt]_\chi$
has something to do with that of QCD and then deduce the chiral
doubling of heavy-light hadrons. This strategy seems to work
amazingly well, giving credence to the notion of the ``vector
manifestation (VM)" of chiral symmetry at chiral restoration
introduced by Harada and Yamawaki~\cite{HY:PR}.
\subsection{The Vector Manifestation of Hidden Local Symmetry}
\indent\indent To make the discussion as simple as possible, I
shall take in the light-quark sector all the current quark masses
to be zero, the so-called chiral limit. The experiments that
brought surprises involve the strange quark whose current quark
mass is comparable to the strong interaction scale, namely, the
pion decay constant $F_\pi\sim 93$ MeV, so to make a quantitative
comparison with experiments, one would have to worry about the
explicit breaking of chiral symmetry. However I believe that the
qualitative feature can be captured in the chiral limit.

Now up to the transition point $[pt]_\chi$, that is, in the chiral
symmetry broken phase, the relevant degrees of freedom are
hadrons. In the standard way of doing things, one assumes that the
only relevant degrees of freedom are the pions with other degrees
of freedom such as vector mesons, baryons etc. considered to be
too heavy to be relevant to the chiral phase transition. This is
the picture typically given by the linear sigma model. The key
point in my discussion which departs from the conventional picture
is that not just the pions but also the vector mesons, namely, the
$\rho$ mesons, are quite relevant. Now how does one ``see" this?
One cannot see it if one has a Lagrangian with the pions and
massive vector mesons coupled in the usual way which is consistent
with the global symmetry but not local gauge invariant because of
the vector-meson mass. With such a Lagrangian it is not easy --
although not impossible -- to go toward $[pt]_\chi$: There is no
systematic way to compute loop corrections. In fact, there is no
easy way to see whether the theory without local gauge invariance
breaks down and if so, locate at what point it does so. There is
however a ``trick" to make the theory locate, and work up to, the
break-down point. That is to introduce hidden local symmetry and
make the theory local-gauge invariant~\cite{georgi}. The authors
in Ref.\cite{georgi} call it ``fake" symmetry but it has the
advantage of endowing the vector mesons with a chiral power
counting~\cite{HY:PR}.

To illustrate the idea, consider the chiral $G\equiv SU(3)_L\times
SU(3)_R$ symmetry appropriate to three-flavor QCD. The symmetry
$G$ is spontaneously broken in the vacuum to $H\equiv
SU(3)_{L+R}$, so the coordinates of the system are given by the
coset space $G/H$ parameterized by the Sugawara field
$U=e^{i\pi/f}$ where $\pi$ is the Nambu-Goldstone pion field. In
the absence of other fields than pions, we have the well-known
chiral perturbation development \`a la Gasser and Leutwyler. In
this pion-only chiral perturbation theory, the vector mesons
$\rho$ can be introduced in consistency with the assumed symmetry.
In fact there are several different ways of doing this but they
are all physically equivalent, provided they are limited to tree
order or the next-to-leading order in chiral perturbation. See
\cite{HY:PR} for a clear discussion on this point. The massive
vectors so introduced do not, however, render themselves to a
systematic chiral perturbation treatment beyond the tree order.
This means that such a theory is powerless as one moves toward the
$[pt]_\chi$ point. In my opinion, works purporting to describe
chiral properties of hot/dense matter away from the vacuum without
resorting to this strategy all suffer from this defect and cannot
be trusted. This difficulty is beautifully circumvented if the
nonlinearly realized chiral symmetry $G/H$ is gauged to linear
$G_{global}\times H_{local}$ as recently worked out by Harada and
Yamawaki. If one fixes the gauge to unitary gauge, one then
recovers the same theory without gauge invariance.

Harada and Yamawaki~\cite{HY:PR} have shown that in hidden local
symmetry theory that exploits the above strategy with pions and
$\rho$ mesons as the relevant degrees of freedom and where a
consistent chiral perturbation can be effectuated, the vector
mesons are found to play an {\it essential role}\ at $[pt]_\chi$
since the mass of the vector meson mass goes to zero in proportion
to the quark condensate $\la\bar{q}q\ra$. In fact by matching the
HLS theory to QCD at a matching scale $\Lambda_M$ above the vector
meson mass, they show by one-loop renormalization-group equation
involving the vector mesons as well pions that $[pt]_\chi$
corresponds to the vector manifestation (VM) fixed point at which
the local gauge coupling $g$ goes to zero and the ratio $a\equiv
F_\sigma^2/F_\pi^2$ (where $\sigma$ are the scalar Goldstone
bosons arising from the spontaneous breaking of the gauge
symmetry) goes to 1.

Up to date, there are no proofs -- lattice or otherwise -- for or
against that the vector mass goes to zero at $[pt]_\chi$. In many
conference talks (e.g., QM2004), one frequently sees view-graphs
in which the $\rho$ and $a_1$ masses come together at the critical
temperature $T_c$ but at non-zero value above the degenerate $\pi$
and $\sigma$. A recent study of chemical equilibration in RHIC
experiments shows that this is most probably
incorrect~\cite{BLR04} in hot matter: Both the $\pi-\sigma$
complex and the $\rho-a_1$ complex should become massless at
$T_c$. Real-time lattice calculations in temperature should
ultimately be able to validate or invalidate this scenario: The
screening mass measured on lattice in hot matter does not carry
the relevant information. In the absence of evidence either for or
against it, we will simply assume that the VM is realized at the
chiral transition point and see whether the result we obtain gives
an a posteriori consistency check, if not a proof, of the
assumption.

The presence of the VM at the chiral transition point $[pt]_\chi$
implies a scenario that is quite different from the standard one
based on the linear sigma model invoked to describe two-flavor
chiral restoration. For instance, the pion velocity at $[pt]_\chi$
is predicted to be near the velocity of light with the vector
mesons at VM~\cite{vpion} whereas the linear sigma model predicts
it to be zero~\cite{ss}.
\subsection{From the VM Fixed Point to the Nambu-Goldstone Phase}
\indent\indent Consider the heavy-light-quark, $Q\bar{q}$, mesons
where $Q$ is the heavy quark and $q$ is the light quark. Again for
simplicity, I shall take the mass of $Q$ to be infinite -- and as
mentioned, that of $q$ to be zero. Let us imagine that we are at
the VM fixed point. In constructing the Lagrangian for the
light-quark system, the relevant variables are the HLS 1-forms for
the light mesons,
 \bea
{\alpha}_{R(L)\mu} &=& \frac{1}{i}
    \partial_\mu \xi_{\rm R(L)} \cdot \xi_{\rm [R(L)]}^\dag
 \eea
which transform under $\chisu3$ as ${\alpha}_{R(L)\mu}\rightarrow
R(L)\,{\alpha}_{R(L)\mu}[R(L)]^\dagger$ with $R(L)\in
SU(3)_R(SU(3)_L)$. Since the gauge coupling $g$ is zero at the
fixed point, the HLS gauge bosons are massless and their
transverse components decouple from the system. Two matrix valued
variables $\xi_{L,R}$ are parameterized as $\xi_{L,R}=\exp
[i\phi_{L,R}]$. Here the combination $(\phi_R + \phi_L)/2$
corresponds to the longitudinal components of the vector mesons
$\rho$ (the $\rho$ meson and its flavor partners) in the broken
phase, while the combination $(\phi_R - \phi_L)/2$ corresponds to
the pseudoscalar Nambu-Goldstone bosons $\pi$ (the pion and its
flavor partners). With these 1-forms and since
$a=(F_\sigma/F_\pi)^2=1$, the light-quark HLS at the VM takes the
simple form,
 \bea
{\mathcal L}_{\rm light}^*=\frac 12 F_\pi^2{\tr}
[\alpha_{R\mu}\alpha_R^\mu +
\alpha_{L\mu}\alpha_L^\mu],\label{lightL}
 \eea
with the star representing the VM fixed point and $F_\pi$ denoting
the bare pion decay constant. The physical pion decay constant
$f_\pi$ vanishes at the VM fixed point by the quadratic divergence
although the bare one is non-zero~\cite{HY:PR}.

For the heavy mesons, one introduces the right and left
fluctuation fields $\calH_R$ and $\calH_L$ that transform under
$\chisu3$ as $\calH_{R(L)}\rightarrow
\calH_{R(L)}R^\dagger(L^\dagger)$. The fixed point Lagrangian for
the heavy mesons in the presence of the light mesons takes the
form
 \bea
{\calL}^*_{\rm heavy}&=& - {\tr}\left[\calH_R
iv_\mu\del^\mu\bar{\calH}_R\right] -{\tr}\left[\calH_L
iv_\mu\del^\mu\bar{\calH}_L\right]+
m_0{\tr}\left[\calH_R\bar{\calH}_R +
\calH_L\bar{\calH}_L\right]\nonumber\\
&&+ 2k\,{\tr}\left[\calH_R{\alpha}_{R\mu}\gamma^\mu
\frac{1+\gamma_5}{2}\bar{\calH}_R  +
\calH_L{\alpha}_{L\mu}\gamma^\mu
\frac{1-\gamma_5}{2}\bar{\calH}_L\right]\,, \label{heavy}
 \eea
where $v_\mu$ is the velocity of the heavy meson, $m_0$ represents
the mass generated by the interaction between heavy quark and the
``pion cloud" surrounding the heavy quark, and $k$ is a real
constant to be determined.

Next we need to consider the modification to the VM Lagrangian
generated by the spontaneous breaking of chiral symmetry. The
gauge coupling constant becomes non-zero, $g \neq 0$,
so the derivatives in the HLS 1-forms become the covariant
derivatives. Then $\alpha_{L\mu}$ and $\alpha_{R\mu}$ are
covariantized:
\begin{eqnarray}
 &\partial_\mu \to D_\mu = \partial_\mu - ig\rho_\mu,&
\nonumber\\
 &\alpha_{R\mu} \to \hat{\alpha}_{R\mu},
\quad
  \alpha_{L\mu} \to \hat{\alpha}_{L\mu}.&
\label{covariant}
\end{eqnarray}
These 1-forms transform as $\hat{\alpha}_{R(L)\mu} \to h\,
\hat{\alpha}_{R(L)\mu}h^\dagger$ with $h \in [SU(3)_V]_{\rm
local}$. Apart from the kinetic-energy term ${\mathcal L}_{\rho
{\rm kin}} = - \frac{1}{2} \mbox{tr} \left[
  \rho_{\mu\nu} \rho^{\mu\nu}
\right]$, there may be other terms, such as e.g., $(a-1)F_\pi^2
\mbox{tr}[\hat{\alpha}_{L\mu} \hat{\alpha}_R^\mu]$ which vanishes
at the fixed point with $a=1$. Although generally $a \neq 1$ in
the broken phase, $a=1$ gives a variety of physical quantities in
good agreement with experiment in matter-free space, as shown in
\cite{HY:PR}. A detailed analysis in preparation for
publication~\cite{HS03} shows that in the present problem, at the
one-loop level that we consider, there are no $(a-1)$ corrections.
Therefore we can safely set $a=1$ in what follows. In the heavy
sector, chiral-symmetry breaking will generate the term
\begin{eqnarray}
{\mathcal L}_{\chi{\rm{SB}}} = \frac{1}{2} \Delta M
\,\mbox{tr}\left[
  \calH_{L} \bar{\calH}_R + \calH_R \bar{\calH}_L
\right] \ , \label{bare Delta M}
\end{eqnarray}
with $\calH_{R(L)}$ transforming under the HLS as $\calH_{R(L)}
\to \calH_{R(L)}h^\dagger$. Here $\Delta M$ is the $bare$
parameter corresponding to the mass splitting between the two
multiplets and can be determined by matching the EFT with QCD.

The main finding of this approach is that $\Delta M$ comes out to
be proportional to the quark condensate, i.e., $\Delta M \sim
\langle \bar{q} q \rangle$.

In order to compute the mass splitting between $D$ and
$\tilde{D}$~\footnote{Although we are referring specifically to
the $D$ mesons, our discussion generically applies to all
heavy-light mesons.}, we need to go to the corresponding fields in
parity eigenstate, $H$ (odd-parity) and $G$ (even-parity) as
defined, e.g., in \cite{NRZ03};
\begin{eqnarray}
&&
 \calH_{R,L} = \frac{1}{\sqrt{2}}\bigl[ G \mp iH\gamma_5 \bigr].
\end{eqnarray}
We shall denote the corresponding masses as $M_{H,G}$. They are
given by
\begin{eqnarray}
M_H &=& - m_0 - \frac{1}{2} \Delta M \ , \nonumber\\
M_G &=& - m_0 + \frac{1}{2} \Delta M \ .
\end{eqnarray}
The mass splitting between $G$ and $H$ is therefore given by
$\Delta M$:
\begin{equation}
M_G - M_H = \Delta M \ .
\end{equation}

The next step in the calculation is to determine $\Delta M$ at the
matching point in terms of QCD variables. We shall do this for the
pseudoscalar and scalar correlators for ${D}(0^-)$ and
$\tilde{D}(0^+)$, respectively. The axial-vector and vector
current correlators can similarly be analyzed for $D(1^-)$ and
$\tilde{D}(1^+)$. In the EFT sector, the correlators are expressed
as
\begin{eqnarray}
 &&
 G_P(Q^2) = \frac{F_D^2 M_D^4}{M_D^2 + Q^2},
\nonumber\\
 &&
 G_S(Q^2) = \frac{F_{\tilde{D}}^2 M_{\tilde{D}}^4}
  {M_{\tilde{D}}^2 + Q^2},
\end{eqnarray}
where $F_D$ ($F_{\tilde{D}}$) denotes the $D$-meson
($\tilde{D}$-meson) decay constant and the space-like momentum
$Q^2 = (M_{D} + \Lambda)^2$ with $\Lambda$ being the matching
scale. If we ignore the difference between $F_D$ and
$F_{\tilde{D}}$ which can be justified by the QCD sum rule
analysis~\cite{Narison:1988ep}, then we get
 \bea
\Delta_{SP}(Q^2) \equiv G_S(Q^2) - G_P(Q^2)\simeq \frac{3 F_D^2
M_D^3}{M_D^2 + Q^2}
   \Delta M_D.
\label{diff-eft}
 \eea
In the QCD sector, the correlators $G_S$ and $G_P$ are given by
the operator product expansion (OPE) as~\cite{Narison:1988ep}
\begin{eqnarray}
  G_S(Q^2) &=&
  \left. G(Q^2) \right\vert_{\rm pert}\nonumber\\
   &+&
  \frac{m_H^2}{m_H^2 + Q^2}
  \Biggl[ {}- m_H \langle \bar{q}q \rangle +
   \frac{\alpha_s}{12\pi}\langle G^{\mu\nu}G_{\mu\nu} \rangle \Biggr],
\nonumber\\
  G_P(Q^2) &=&
  \left. G(Q^2) \right\vert_{\rm pert}\nonumber\\
   &+&
  \frac{m_H^2}{m_H^2 + Q^2}
  \Biggl[ m_H \langle \bar{q}q \rangle +
   \frac{\alpha_s}{12\pi}\langle G^{\mu\nu}G_{\mu\nu} \rangle \Biggr],
\end{eqnarray}
where $m_H$ is the heavy-quark mass. To the accuracy we are aiming
at, the OPE can be truncated at ${\mathcal O}(1/Q^2)$. The
explicit expression for the perturbative contribution $\left.
G(Q^2) \right\vert_{\rm pert}$ is available in the literature but
we do not need it since it drops out in the difference. {}From
these correlators, the $\Delta_{SP}$ becomes
\begin{equation}
 \Delta_{SP}(Q^2) = - \frac{2 m_H^3}{m_H^2 + Q^2}
  \langle \bar{q}q \rangle.
\label{diff-ope}
\end{equation}
Equating Eq.~(\ref{diff-eft}) to Eq.~(\ref{diff-ope}) and
neglecting the difference $(m_H-M_D)$, we obtain the following
matching condition:
\begin{equation}
 3 F_D^2 \, \Delta M_D \simeq - 2 \langle \bar{q}q \rangle.
\label{bare mass diff}
\end{equation}
Thus at the matching scale, the splitting is
\begin{equation}
 \Delta M_D \simeq -\frac{2}{3}\frac{\langle \bar{q}q \rangle}{F_D^2}.
 \label{bare mass diff matching}
\end{equation}
As announced, the splitting is indeed proportional to the
light-quark condensate. Let us denote the $\Delta M_D$ determined
at the scale $\Lambda$ as $\Delta M_{\rm bare}$ which will figure
in the numerical calculation.

\begin{figure}[ht]
\centerline{\epsfxsize=6cm\epsfbox{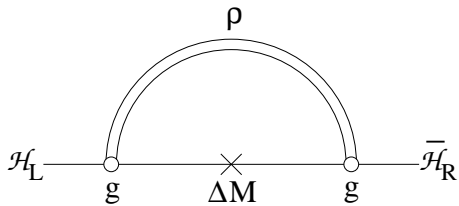}} \caption{Diagram
contributing to the mass difference.}\label{fig:mass}
\end{figure}

Given the splitting $\Delta M_{\rm bare}$ at the scale $\Lambda$,
we need to decimate down to the physical scale. This amounts to
making quantum corrections to the correlators written in terms of
the bare quantities or more specifically to ${\mathcal
L}_{\chi{\rm{SB}}}$ in Eq.~(\ref{bare Delta M}). This calculation
turns out to be surprisingly simple for $a= 1$. For $a=1$,
$\phi_L$ does not mix with $\phi_R$ in the light sector, and hence
$\phi_L$ couples to only $\calH_L$ and $\phi_R$ to only $\calH_R$.
As a result $\calH_{L(R)}$ cannot connect to $\calH_{R(L)}$ by the
exchange of $\phi_L$ or $\phi_R$. Only the $\rho$-loop links
between the fields with different chiralities as shown in
Fig.~\ref{fig:mass}.

This term contributes to the two-point function as
\begin{eqnarray}
 \biggl. \Pi_{LR} \biggr\vert_{\rm div}
 = - \frac{1}{2} \Delta M \,
   {\mathcal C}_2(N_f) \,
   \frac{g^2}{2\pi^2}\Bigl( 1 - 2k - k^2 \Bigr)
   \ln\Lambda,
\label{quantum correction}
\end{eqnarray}
where ${\mathcal C}_2(N_f)$ is the second Casimir defined by
$(T_a)_{ij}(T_a)_{jl}= {\mathcal C}_2(N_f)\delta_{il}$ with $i, j$
and $l$ denoting the flavor indices of the light quarks. This
divergence is renormalized by the bare contribution of the form
$\Pi_{LR,{\rm bare}} = \frac{1}{2}\Delta M_{\rm bare}$. Thus the
renormalization-group equation (RGE) takes the form
\begin{equation}
 \mu\frac{d\,\Delta M}{d\mu}
 =  {\mathcal C}_2(N_f)\,
   \frac{g^2}{2\pi^2}\Bigl( 1 - 2k - k^2 \Bigr)\Delta M.
\label{rge}
\end{equation}
For simplicity, we may neglect the scale dependence in $g$ and
$k$. Then the solution to the RGE for $\Delta M$ is
\begin{equation}
 \Delta M = \Delta M_{{\rm bare}}
 \exp\Bigl[ - {\mathcal C}_2(N_f)\,
   \frac{g^2}{2\pi^2}\Bigl( 1 - 2k - k^2 \Bigr)
   \ln\frac{\Lambda}{\mu} \Bigr].
\label{mass diff}
\end{equation}
This is our main result. This shows unequivocally that the mass
splitting is dictated by the $bare$ splitting $\Delta M_{\rm
bare}$ proportional to $\la\bar{q}q\ra$ corrected by the quantum
effect given by $C_{\rm quantum}=\exp\Bigl[ - {\mathcal
C}_2(N_f)\,
   \frac{g^2}{2\pi^2}\Bigl( 1 - 2k - k^2 \Bigr)
   \ln\frac{\Lambda}{\mu} \Bigr]$.
\section{Prediction}
\subsection{$\Delta M$}
\indent\indent In the chiral limit, one can make a neat prediction
on the splitting $\Delta M$. There are no free parameters here.

I shall not attempt any error analysis and merely quote the
semi-quantitative estimate arrived at in \cite{HRS-babar}. The
second Casimir for three flavors is ${\mathcal C}_2(N_f=3)=4/3$;
the constant $k$ can be extracted from $D^*\rightarrow D\pi$
decay~\cite{Hagiwara:fs} and comes out to be $k \simeq 0.59$. By
taking $\mu = m_\rho = 771\,\mbox{MeV}$, $\Lambda =
1.1\,\mbox{GeV}$ and $g = g(m_\rho) = 6.27$ determined through the
Wilsonian matching~\cite{HY:PR}, we find that the quantum effect
increases the mass splitting by about 60\%, i.e., $C_{\rm
quantum}\approx 1.6$. It turns out~\cite{HS03} that this result is
quite stable against the matching scale $\Lambda$. Taking the
value for $F_D$, $F_D \simeq 0.205$ GeV, and that for $\langle
\bar{q}q \rangle$, $\langle \bar{q}q \rangle = -(0.243
\,\mbox{GeV})^3$ from the literature~\cite{Narison:2003td} as
typical ones, we find from (\ref{bare mass diff matching})
 \bea
\Delta M_{\rm bare}\simeq 0.23 \ {\rm GeV}
 \eea
so that
 \bea
\Delta M\simeq 0.37 \ {\rm GeV}. \label{mass splitting}
 \eea
This should be compared with the constituent quark mass $\simeq
m_N/3$ where $m_N$ is the nucleon mass. This is consist with what
was observed in the experiments~\cite{BABAR,CLEO}. Of course, in
comparing with experiments, particularly the BaBar/CLEO
experiments, we need to take into account the flavor symmetry
breaking which is not yet systematically investigated in the
framework discussed here. But the point is that it is the quark
condensate that carries the main imprint of the splitting. Another
point of interest in the result is that the bare splitting depends
on the heavy-meson decay constant. This suggests that the
splitting may show heavy-quark flavor dependence. This could be
checked with experiments once a systematic heavy-quark expansion
(which is not done here) is carried out.
\subsection{Implications}
\indent\indent There is an obvious implication on heavy-light
baryons that can be obtained as skyrmions~\cite{RRS,jmw} from the
heavy-light mesonic Lagrangian. One expects off-hand that the
chiral doubling splitting in heavy-light baryons would also be
given by the $\rho$-exchange graph and hence will likewise be
proportional to the light-quark condensate. Another exciting
avenue would be to look at pentaquarks as skyrmions in this
HLS/VM-implemented theory with a heavy quark replacing the strange
quark in the recently observed $\Theta^+$ baryon which is
generating lots of activities nowadays. It would be interesting to
expose the contribution to the heavy penaquark mass that bears
directly on chiral symmetry as in the heavy-light mesons.

Suppose future experiments do show that in hot/dense matter, the
splitting in heavy-light mesons or baryons gets reduced as
temperature/density goes up in such a way as to be consistent with
the vanishing splitting at the critical point in the chiral limit.
An attractive interpretation of such an observation is that one is
realizing the VM at $[pt]_\chi$, and hence the $\rho$ meson mass
does go to zero at the phase transition as predicted in a
different context a long time ago~\cite{BR91}. Furthermore a
recent striking development~\cite{BLR04} on the phase structure of
hot matter near $T_c$ suggests that massive excitations in the
$\rho$ channel above $T_c$ in the form of an instanton liquid go
massless at $T_c$ as do the pion and the scalar $\sigma$. Lattice
confirmation of this phenomenon would be highly desirable.
\section{Concluding Remarks}
\indent\indent This is a ``concluding" talk in more than one
sense. It is the last talk in this Symposium and is also most
likely the last talk in this series of astro-hadron physics I have
been helping develop in KIAS. So let me add a few of my personal
remarks here.

In early 1990's, with a small group of young -- as well as less
young -- theorists in hadronic physics in Korea I initiated a
concerted effort to understand how hadronic physics involved in
the strong interactions of matter can be merged into certain
aspect of astrophysical phenomena that are thought to be produced
under extreme conditions of temperature and/or density, a new
field of research which we called ``astro-hadron physics." The
first international meeting in Korea bearing that name -- funded
by APCTP -- was held at Seoul National University in 1997. With
the advent of Korea Institute for Advanced Study (KIAS) originally
conceived with the primary purpose of generating and developing
original, innovative research activities in Korea that could be
brought to the forefront of the world, the activity in
astro-hadron physics was taken up at KIAS in the precise spirit of
the institute's objective. With the influx of a large number of
bright visitors from abroad, the activity has met with success.
This then led to the first KIAS astrophysics meeting in 2000 in
which astro-hadron physics figured importantly in bringing
together such explosive astrophysical processes as supernovae,
gamma-ray burst, black-hole formations with such explosive
laboratory processes as relativistic heavy-ion collisions. The
so-called hadronic phase diagram shown at this meeting was quite
barren with most of the areas unexplored or empty, with little
overlap between what the astronomers were observing and what the
laboratory experimenters were measuring. Since then, the phase map
has rapidly filled up, as we witnessed in this meeting, with
measurements coming from various terrestrial laboratories (CERN,
RHIC...) and from satellite observatories (Chandra, RXTE ...).
This meeting is clearly a timely one to start establishing crucial
connections between the two sources and synthesizing a coherent
picture that will ultimately expose the structure of the novel
form of matter searched for in extreme conditions of temperature
and/or density.

Although this may be -- at least for the time being -- the last
meeting of the series here at KIAS, the activity in this field
should, and surely will, go on, if not here, then elsewhere in
this country. With the advent of JPARC in Japan in tandem with
that of SIS 300 at GSI in Germany together with forthcoming
satellite observatories, this field is poised to develop strongly
in this Asian Pacific area. It would be a pity if Korea with her
early start were to miss out in this exciting new development.
What I discussed in my talk together with the discovery of the
novel structure in pentaquark systems promise clearly that there
will be surprises and breakthroughs in store in this field.

\subsection*{Acknowledgments}
\indent\indent I would like to thank Masayasu Harada, Maciek
Nowak, Chihiro Sasaki and Ismail Zahed for most enjoyable
collaborations on which my talk was based.

\def\bi{\bibitem}

\end{document}